\documentclass[lettersize,journal]{IEEEtran}
\usepackage{amsmath,amsfonts}
\usepackage{algorithmic}
\usepackage{algorithm}
\usepackage{array}
\usepackage[caption=false,font=normalsize,labelfont=sf,textfont=sf]{subfig}
\usepackage{textcomp}
\usepackage{stfloats}
\usepackage{url}
\usepackage{verbatim}
\usepackage{graphicx}
\usepackage{cite}
\usepackage{tabularx} 
\usepackage{makecell} 
\usepackage{textcomp}
\usepackage{booktabs}
\usepackage{color}
\usepackage{algorithm}
\usepackage{algorithmic}
\usepackage{float}
\usepackage{multirow}

\begin{document}

\title{Road to 6G Digital Twin Networks: Multi-Task Adaptive Ray-Tracing as a Key Enabler}

\author{Li Yu, Yinghe Miao, Jianhua Zhang, Shaoyi Liu, Yuxiang Zhang, and Guangyi Liu}

\maketitle

\begin{abstract}

As a virtual, synchronized replica of physical network, the digital twin network (DTN) is envisioned to sense, predict, optimize and manage the intricate wireless technologies and architectures brought by 6G. Given that the properties of wireless channel fundamentally determine the system performances from the physical layer to network layer, it is a critical prerequisite that the invisible wireless channel in physical world be accurately and efficiently twinned. To support 6G DTN, this paper first proposes a multi-task adaptive ray-tracing platform for 6G (MART-6G) to generate the channel with 6G features, specially designed for DTN online real-time and offline high-accurate tasks. Specifically, the MART-6G platform comprises three core modules, i.e., environment twin module to enhance the sensing ability of dynamic environment; RT engine module to incorporate the main algorithms of propagations, accelerations, calibrations, 6G-specific new features; and channel twin module to generate channel multipath, parameters, statistical distributions, and corresponding three-dimensional (3D) environment information. Moreover, MART-6G is tailored for DTN tasks through the adaptive selection of proper sensing methods, antenna and material libraries, propagation models and calibration strategy, etc. To validate MART-6G performance, we present two real-world case studies to demonstrate the accuracy, efficiency and generality in both offline coverage prediction and online real-time channel prediction. Finally, some open issues and challenges are outlined to further support future diverse DTN tasks.

\end{abstract}


\section{Introduction}

The sixth-generation (6G) communication system is expected to offer intelligent, hyper-reliable, and ubiquitous connectivity, supporting the internet of everything (IoE) trend for future digital society\cite{1_6GWCM,6_zhang2023channel}. However, the global coverage and diverse scenarios brought by 6G pose extreme complexity and new challenges for network design and technology optimization. The digital twin (DT) technique is considered an innovative tool to overcome these challenges. By creating a real-time replica of physical entities in the digital world, the digital twin network (DTN) can offer up-to-date network status, closed-loop decisions, and real-time interaction between the digital world and physical world, enabling proactive adaptation to the complex wireless network for 6G\cite{2_DTN2024WirelessCommunications}.

To realize the vision of 6G DTN, accurate wireless channel between the transmitter (Tx) and receiver (Rx) is required, encompassing propagation parameters, path loss (PL), and other relevant attributes\cite{3_wanghengDTC}. It facilitates decision-making for various wireless tasks, such as network planning, resource allocation, and link operation. However, traditional channel modeling primarily relies on statistical analysis of channel measurements, which struggles to accurately describe wireless channel characteristics at specific times and locations, failing to meet the requirements of DTN. Fortunately, deterministic ray-tracing (RT), which relies on the ray-optic approximation, can theoretically simulate the multipath components (MPCs) between any Tx-Rx pair in space with high precision\cite{4_beijiaoRT}.

RT, as an essential tool for constructing high-precision twin channel in the digital world, is expected to become a key enabler of 6G DTN \cite{5_NVIDADTN}. First, RT can be rapidly deployed in new scenarios without relying on a large amount of measurement data. Second, RT demonstrates significant applicability in simulating and evaluating emerging 6G technologies. For example, it can simulate the high-frequency sparsity of terahertz (THz) signals, the sensing and positioning in integrated sensing and communication (ISAC) scenarios, the radio modulation of reconfigurable intelligent surfaces (RIS), and the high spatial resolution demands of extremely large-scale multi-input multi-output (XL-MIMO). Third, RT relies heavily on high-precision digital maps and substantial computational resources. Thanks to recent advances in map reconstruction and graphics processing units (GPUs), RT offers new opportunities to advance DTN development.

\begin{table*}[h]
	\centering
	
	\caption{Comparison of Commercial and public Academic RT Simulators for 6G.}
	\begin{tabular}{c|c|c|c|c|c|@{}c@{}}
		
		\hline
		\textbf{Simulator} & \textbf{Frequency} & \textbf{Propagation mechanisms} & \textbf{\makecell{6G\\ technologies}} & \textbf{\makecell{Computational\\acceleration}} & \textbf{\makecell{Task-oriented \\ customization}} & \textbf{Calibration} \\ \hline 
		\makecell{MATLAB ray\\ tracing toolbox\\ \cite{7_matlab_ray_tracing}} & \makecell{100 MHz-\\100 GHz} & \makecell{Direct, reflection, \\diffraction,\\ transmission \\(need to customize materials)} & \makecell{V2X} & \makecell{CPU multi-thread\\ acceleration} & No & N/A \\ \hline
		\makecell{Volcano\cite{8_siradel_volcano}} & \makecell{Sub-6G \\to mmWave} & \makecell{Direct, reflection,\\ diffraction, transmission} & \makecell{XL-MIMO} & \makecell{GPU and parallel\\ computing} & \makecell{No} & \makecell{Manual calibration} \\ \hline
		\makecell{Wireless\\ InSite\cite{9_remcom_wireless_insite}} & \makecell{50 MHz\\ to THz} & \makecell{Direct, reflection\\ diffraction, transmission,\\ scattering} & \makecell{THz, \\ XL-MIMO, RIS} & \makecell{GPU and parallel\\ computing,\\ simplification of \\complex geometric \\models} & No & \makecell{Manual calibration} \\ \hline
        \makecell{NYURay\cite{10_NYURay}\\(academic)} & \makecell{mmWave \\to THz} & \makecell{Direct, reflection,\\ diffraction, transmission, \\scattering} & \makecell{THz, XL-MIMO} & \makecell{Dimension reduction,\\ GPU and parallel \\computing} & \makecell{No}&\makecell{Extensive manual\\ calibration} \\ \hline
		\makecell{Cloud RT\cite{4_beijiaoRT}\\(academic)} & \makecell{400 MHz\\-350 GHz} & \makecell{Direct, reflection, \\diffraction, transmission} & \makecell{XL-MIMO, V2X} & \makecell{Space partition,\\ multi-thread and \\distributed computing} & \makecell{No} & \makecell{Several published\\ calibration results} \\ \hline
		\makecell{Sionna\cite{11_sionna}\\(academic)} & \makecell{0.5-100 GHz} & \makecell{Direct, reflection,\\ diffraction (simplified model), \\transmission} & \makecell{XL-MIMO, RIS} & \makecell{GPU and parallel\\ computing} & No & \makecell{Gradient descent \\automatic calibration } \\ \hline
		\makecell{6G adaptive \\ ray-tracing\\(academic)} & \makecell{400 MHz\\ to THz} & \makecell{Direct, reflection, \\diffraction, transmission,\\ scattering} & \makecell{THz, ISAC,\\ XL-MIMO,\\ RIS, V2X} & \makecell{Environment \\prior SBR,\\Dynamic RT,\\ GPU and parallel\\ computing} & Yes & \makecell{Machine learning \\automatic calibration} \\ \hline		
	\end{tabular}
	\label{table_RT}
\end{table*}

Up to now, extensive research on RT for 6G has been conducted, with multiple advanced RT simulators developed by academia and industry. Mainstream commercial software including Volcano\cite{8_siradel_volcano} and WirelessInSite\cite{9_remcom_wireless_insite} supports simulations across the Sub-6G to mmWave frequency bands and includes new technology like XL-MIMO. Notably, WirelessInSite's latest version implements RIS modeling. Academic simulators prioritize model accuracy. Sionna\cite{11_sionna} is a differentiable RT simulator that supports gradient descent-based self-calibration. NYURay\cite{10_NYURay} has made significant contributions with its extensive calibration work in the THz band. Most simulators feature hardware acceleration based on central processing units (CPUs) and GPUs, and Cloud RT [4] enables algorithmic optimizations like spatial partitioning. However, most simulators do not enable simulations for multiple new technologies. Furthermore, DTN presents new challenges for RT, particularly balancing model accuracy and complexity according to communication task requirements. The absence of task-specific adaptation mechanisms in existing simulators may hinder their applicability to diverse DTN scenarios.

Therefore, a RT simulator that supports diverse 6G technologies and enables DTN task customization is required. In this paper, we propose a multi-task adaptive ray-tracing platform for 6G (MART-6G), comprising three core modules, i.e., environment twin, RT engine and channel twin, specially designed for DTN tasks. MART-6G is customized to both online and offline DTN tasks to generate channel parameters through the selection of proper antenna and material libraries, propagation mechanisms, acceleration algorithms and 6G-specific features. Additionally, accurate channel models for 6G technologies, such as THz, XL-MIMO, ISAC, and RIS, are integrated to empower DTN in assessing 6G transmission performance. We believe that MART-6G enhances DTN's capability to accurately characterize and efficiently exploit propagation environments. Therefore, integrating MART-6G into DTN is envisioned to help achieve an autonomous and intelligent 6G network, which features proactive adaptation to environmental changes and optimal decision-making for communication tasks.

\section{RT for 6G}
RT is a deterministic channel modeling method based on geometric optics (GO) and uniform theory of diffraction (UTD). It tracks electromagnetic wave propagation paths, including reflection, transmission, diffraction, and scattering, to accurately simulate multipath propagation. This aligns with 6G requirements but also introduces new challenges. This section summarizes 6G channel research demands and examines RT's challenges and advancements for 6G.

\subsection{New Demand for 6G Channel}
\textbf{New Band Requirements:} As 6G expands into the terahertz and mid-high frequency bands (7–24 GHz) introduces challenges. The terahertz band offers Tbps-level bandwidth but faces high PL, atmospheric absorption, and complex scattering. The mid-high band frequency balances coverage and capacity but struggles with weak diffraction, high penetration loss, and dynamic blockage. Accurate modeling of reflection, transmission, and scattering is crucial for 6G development.

\textbf{New Technology Requirements:}  The introduction of key 6G technologies such as ISAC, XL-MIMO, and RIS has posed revolutionary demands on the dimensionality and accuracy of channel modeling. ISAC requires embedding of target scattering characteristics into channel models, XL-MIMO requires spatial non-stationarity and near-field effects, and RIS calls for modeling phase reconfiguration's impact on propagation. To address these demands, high-precision channel modeling methods for key 6G technologies must be developed.

\textbf{New Scenario Requirements:} 6G expands into low-altitude, high-speed, and industrial scenarios, demanding adaptive channel models. Low-altitude communication needs modeling of drone trajectories, building diffraction, and ground reflection, with meteorological impacts on mmWave/THz links. Smart factories require models for multipath interference, robotic movements, and electromagnetic compatibility. These challenges necessitate integrating the high-precision 3D digital twins with real-time sensing data to dynamically reconstruct channel characteristics for various industries.

\subsection{Challenges and Progress for 6G RT}
The core of RT lies in its propagation modeling driven by physical mechanisms. This modeling can integrate the geometric information of the environment and the electromagnetic parameters of materials to predict multi-dimensional features, such as delay, angle, and polarization in the channel impulse response (CIR). Due to its high accuracy, interpretability, and generalizability across scenarios, RT is expected to play a crucial role in 6G systems. However, its practical implementation still faces three main challenges: (1) Imperfect propagation characteristic modeling for new frequency bands, including atmospheric absorption attenuation and complex scattering mechanisms in THz bands; (2) Critical gaps in electromagnetic response modeling for emerging technologies, particularly scattering target characterization for ISAC, cascaded channel reconstruction for RIS, and near-field coupling for XL-MIMO; (3) Insufficient task-oriented customization, requiring further research into adaptive RT algorithms and acceleration strategies tailored to different wireless task requirements, to balance model accuracy and complexity.

In response to the aforementioned challenges, both academia and industry have proposed various solutions. This paper summarizes the features of the existing mainstream RT simulators, as shown in Table \ref{table_RT}. Currently, mainstream RT simulators provide insufficient support for new 6G technologies. Among them, Wireless InSite offers the most comprehensive support, including simulations for THz, XL-MIMO, and RIS \cite{9_remcom_wireless_insite}. However, it shows limitations in dynamic scenario simulations. In academic-oriented RT simulators, some support for 6G technologies is present. For instance, NYURay supports THz and XL-MIMO, while Sionna supports XL-MIMO and RIS \cite{10_NYURay,11_sionna}. However, no RT simulator above currently supports a comprehensive range of 6G technologies, nor do they offer adaptive strategies for different communication tasks. Meanwhile, due to modeling errors and other factors, RT platforms cannot fully fit real-world measurement data, requiring efficient calibration to ensure result accuracy. To meet the future needs of the 6G DTN, developing a new RT simulator is still required.

\section{RT-Enabled Digital Twin Network}

\begin{figure*}[htbp]
\centerline{\includegraphics[scale=0.50]{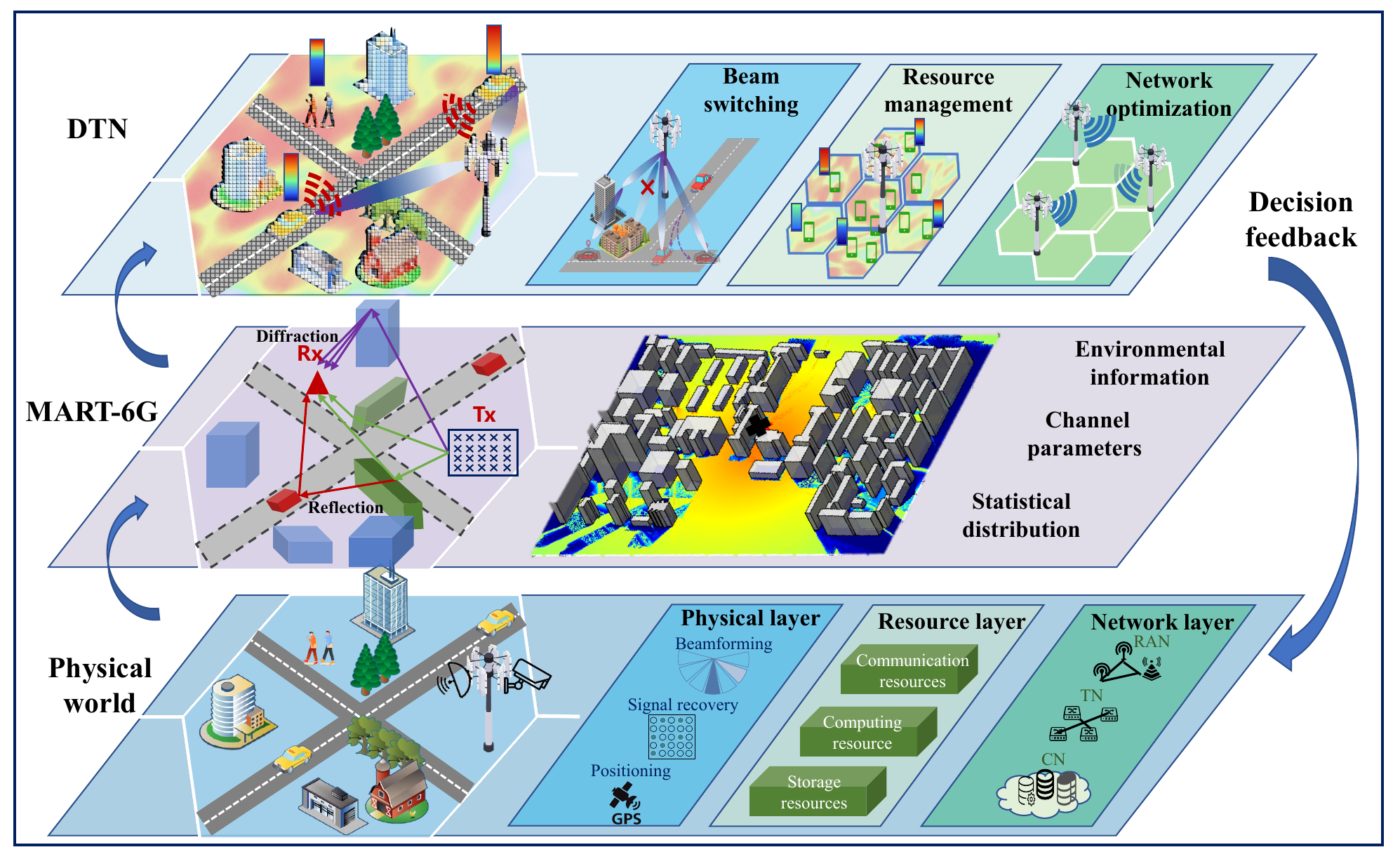}}
\caption{Illustration of RT enabled 6G DTN. }
\label{GUI_RT}
\end{figure*}

In this section, we provide a detailed introduction to RT-enabled DTN. Additionally, we categorize the communication tasks of DTN into online twin tasks and offline twin tasks, each with distinct requirements for RT.

\subsection{DTN Requirements}
DTN serves as a digital replica of the full life cycle of the physical network. It uses data and models to create an accurate network simulation platform, which reflects the current network status and predicts future states. Typically, DTN is divided into twins for the network, resource, and physical layers. The network layer twin models network elements, functions, and application services. The resource layer twin monitors and manages storage, computing, and communication resources. The physical layer twin reconstructs physical entities, links, and transmission technologies. Similar to physical networks, the wireless channel plays a critical role in determining performance metrics such as transmission rates, signal quality, and network capacity. As such, the digital replica of the channel forms the foundation for the construction and operation of the DTN. Each twin layer supports distinct communication tasks, which require different twin channels and real-time performance. Specifically, communication tasks for twin channels are categorized as either offline or online twins, based on their varying real-time and accuracy requirements.

\textbf{Offline Twin Tasks:} These tasks typically focus on the network and resource layers, including network planning, operation, and management. Unlike tasks in the physical layer, which require real-time channel data at the millisecond (ms) level, offline twin tasks rely on updates over extended periods, such as hours or days. Additionally, they primarily involve large-scale channel information, such as received power and signal-to-noise ratio (SNR). For example, in network planning, DTN is used to obtain accurate power coverage across various base station locations, with city-level coverage calculations completed within a few hours deemed acceptable.

\textbf{Online Twin Tasks:} These tasks are typically focused on the physical layer, addressing specific user links, such as channel coding and modulation, beam management, and other related functions. They often require real-time channel data with second-level or ms-level timeliness. Additionally, online twin tasks emphasize small-scale channel information, such as the angular distribution of power and complete channel state information (CSI). For instance, DTN is used for beam tracking of a base station for a mobile user, requiring real-time channel angle distribution data based on the user's current location in order to direct the beam toward the direction with minimal loss. In this case, update intervals on the order of seconds are considered acceptable.

\begin{figure*}[htbp]
\centerline{\includegraphics[scale=0.33]{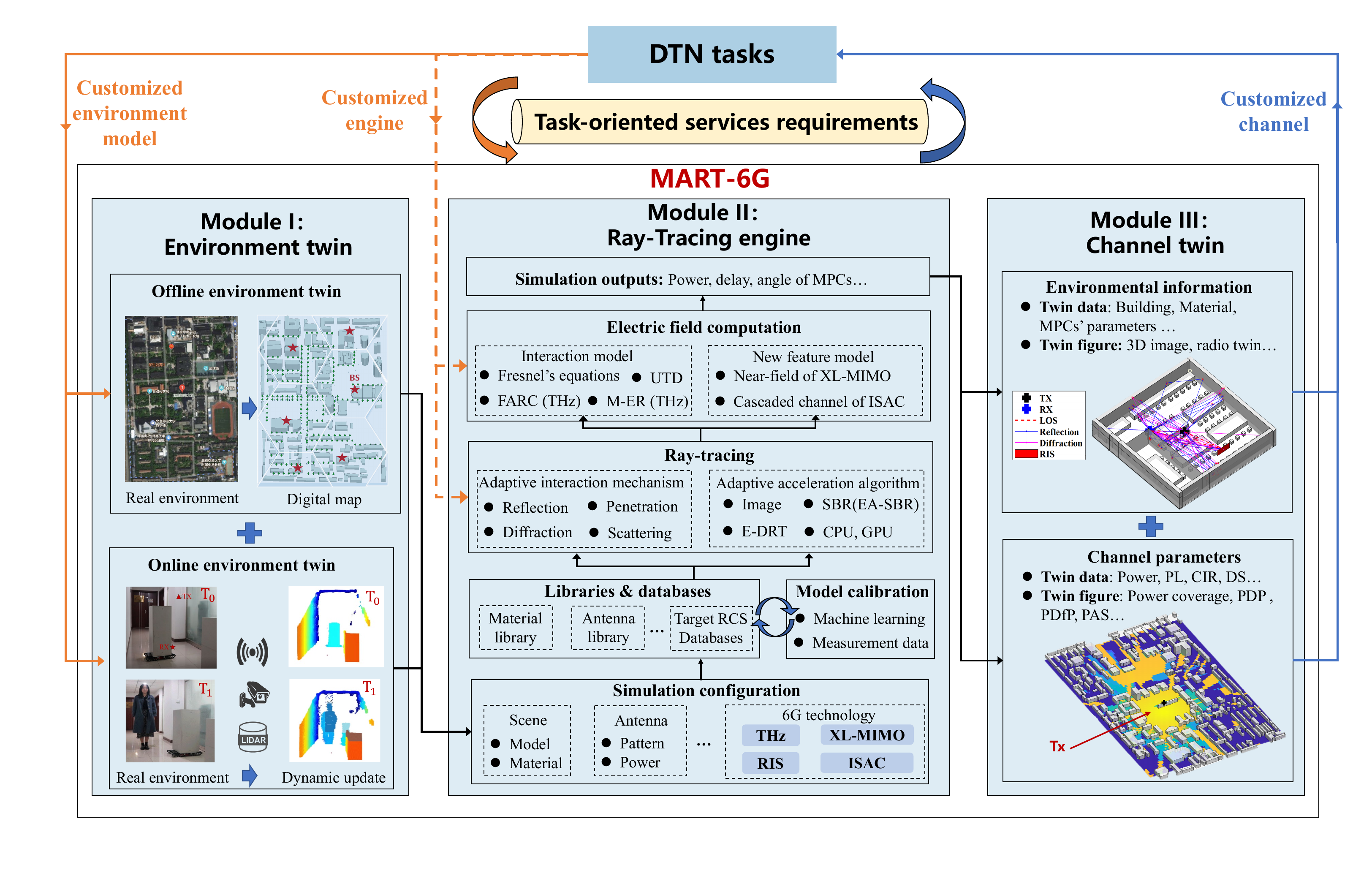}}
\caption{The architecture of the MART-6G, which is a task-oriented customized simulation engine for DTN, also supports the evaluation of potential 6G technologies.}
\label{arc_RT}
\end{figure*}

\subsection{RT as A Key Enabler}
Based on the demands of DTN, the MART-6G is designed, which is responsible for the precise environment twin and the radio propagation twin using data from the physical world. Depending on the characteristics of the tasks issued by DTN, twin channels with varying accuracy and timeliness are constructed. The core of the MART-6G is an intelligent RT engine that can dynamically adjust the environmental model's accuracy, propagation mechanisms, and RT acceleration algorithms in order to balance complexity and precision. The twin channels of the MART-6G are multi-modal, incorporating large-scale data such as power map and PL, as well as small-scale data like the parameters of MPCs and CIR, to meet the needs of various tasks. Additionally, the visualization of twin channels is an essential feature, including the display of current radiomap, SNR, and other parameters, which allows engineers to more intuitively observe changes in channel conditions. Examples of specific tasks are also provided.

\textbf{Network Operation and Management (Offline Twin Tasks):} These tasks aim to create virtual drive tests using DTN prior to actual network configuration changes, ensuring their effectiveness. These tasks require Ray Tracer to provide parameters such as reference signal received power (RSRP) and channel quality indicator (CQI) for virtual drive tests, with no sensitivity to simulation time.

\textbf{Signal Decoding and Recovery (Online Twin Tasks):} Accurate CSI is essential for high-precision signal recovery. In physical networks, CSI is typically obtained through channel estimation based on pilot signals. In DTN, however, CSI can be fully computed by Ray Tracer, which eliminates the need for pilot signals and significantly reduces overhead. This task requires Ray Tracer to balance model complexity and accuracy, ensuring that computation delays remain within milliseconds while achieving at least 90\% CSI accuracy.

\textbf{Radio-based Sensing and Localization (Online Twin Tasks):} In ISAC scenarios, this task uses radio propagation information for target sensing and localization. High-accuracy localization requires the angle and delay parameters of MPCs for radio inversion. The Ray Tracer must ensure that more than 80\% of the energy in the MPCs is accurately replicated, with simulation times kept within the range of seconds.

\section{Architecture and Capabilities of MART-6G}
This section details the architecture and key features of the MART-6G.

\subsection{The Framework and Workflow  }
The MART-6G is designed with three core modules: the environment twin module, the RT engine module, and the channel twin module.

\textbf{Environment Twin Module:} This module is responsible for constructing the twin model of the physical world, where accuracy directly impacts RT performance. Large-scale static twin environments are built using high-precision digital maps (e.g., satellite images, radar point clouds), with geometric errors within centimeters. Additionally, the electromagnetic properties of materials, which influence radio propagation, should also be calibrated through measurements. For online tasks, the module tracks the positions and motion of moving objects like vehicles and pedestrians. With advances in autonomous driving, dynamic twin functionality is now in practical use.

\textbf{RT Engine Module:} This module calculates radio propagation in the twin environment, encompassing phenomena such as reflection, diffraction, penetration, and scattering, to obtain the parameters of MPCs for the radio twin. specific simulation setups are tailored to the DTN task requirements, covering scenarios, antennas, and the 6G technologies to be activated. Then, adaptive RT simulation is executed using the calibrated material library, antenna library, and the radar cross-section (RCS) database of the sensing targets. Specifically, the RT engine dynamically adjusts interaction mechanisms and their order based on empirical data. Different acceleration algorithms such as the environment information-assisted shooting and bouncing rays (EA-SBR)\cite{12_E_SBR}, and enhanced dynamic RT (E-DRT)\cite{13_EDRT} are applied, offering varying acceleration efficiencies and precision guarantees. Thanks to NVIDIA’s light cone acceleration, GPU acceleration boosts speed by over 10 times. Finally, electric fields of rays are calculated using interaction models, such as Fresnel’s equations, UTD, and new models for specific frequency bands. At the same time, new 6G channel characteristics are modeled, including near-field of XL-MIMO and target scattering characteristics for ISAC.

\textbf{Channel Twin Module:} This module is responsible for extracting channel characteristics and constructing the channel twin. It can output comprehensive channel features, including large-scale information such as PL, delay spread (DS), and small-scale information like CIR. Based on the service requirements of different DTN tasks, this module outputs customized channel parameters. Additionally, channel characteristic visualization is an important feature, including the presentation of cell power coverage, and RSRP of mobile users.

\subsection{New Features of MART-6G}

\textbf{Support 6G New Features:} The MART-6G’s key feature is its ability to simulate and evaluate potential 6G technologies, such as THz, XL-MIMO, ISAC, and RIS. For THz simulation, when the carrier frequency exceeds 0.1 GHz, the THz band propagation model is activated. The frequency-angle two-dimensional reflection coefficient (FARC) model\cite{6_zhang2023channel} is used to calculate the reflection. The modified effective roughness (ER) model\cite{14_scatter} is also applied to model random scattering, and the cloud fog attenuation is considered for PL. In XL-MIMO simulation, when the increase in array size causes the propagation distance to exceed the Rayleigh distance, the plane wave assumption is no longer used to synthesize the MIMO channel matrix. Instead, all antenna elements are traced. To enhance efficiency, an efficient RT\cite{15_yuanRTMIMO} algorithm is employed, preserving near-field and non-stationary characteristics while improving efficiency by over 90\%. For ISAC and RIS, Ray Tracer constructs models for the RCS of the sensing targets and the RIS response, simulating their effects on radio. The RCS of targets, such as cars, drones, and pedestrians, is updated according to measurement. Additionally, a RIS response model based on physical optics\cite{6_zhang2023channel} is used, aligning well with measured results.

\textbf{DTN Task-Oriented Customization:} The MART-6G is a customized simulation engine for DTN, designed to dynamically adjust configurations and models to meet the service requirements of different tasks. For offline twin tasks such as network planning, high-precision textured environmental models are enabled, and propagation paths are calculated using IM or SBR algorithms with a high number of emitted rays. The allowed interaction order is typically up to 4, ensuring that power errors remain below 6 dB. For online twin tasks like beam management, the engine balances model accuracy and complexity. Generally, lower-resolution environmental models and lightweight SBR with related acceleration strategies are employed, limiting interactions to below 4. In this case, over 90\% of the power-carrying MPCs are generally detected, and thanks to GPU parallel computing, the simulation time is kept within the ms range.

\textbf{Calibration Based on Measurements:} To ensure the accuracy of the twin channel, the database and models of the MART-6G are calibrated using a large amount of measurement data. On the one hand, the electromagnetic parameters of typical materials across different frequency bands, including dielectric constant and conductivity, are strictly calibrated using machine learning algorithms such as simulated annealing. On the other hand, a self-calibration interface is provided for users, allowing material parameters to be self-calibrated based on user-provided measurement data, such as RSRP.

\section{Case Studies}
In this section, we use the MART-6G to generate the required channel data for both offline and online communication tasks. These experiments demonstrate the advantages and feasibility of the MART-6G.
\subsection{Offline Task Validation}

For the offline validation, we use the network planning task assigned by DTN and select the Beijing University of Posts and Telecommunications campus as the target scenario. The digital map is obtained from OpenStreetMap, with physical base station locations used as Tx. The carrier frequency is 14.8 GHz in new mid-high frequency band, which is expected to become the operational spectrum for 6G. To validate the platform’s accuracy and calibration, as shown in Fig. 3(a), we selected 73 Line-of-Sight (LoS) and 46 Obstructed Line-of-Sight (OLoS) points for channel measurements using omnidirectional antennas. Thirty points were randomly chosen as validation points, and the remaining ones were used for calibration. The simulated annealing algorithm is used to calibrate the material parameters. After calibration, power coverage simulations were performed across the scene at 1-meter intervals to help identify coverage gaps and optimize the network.
\begin{figure}[tbp]
\centerline{\includegraphics[scale=0.25]{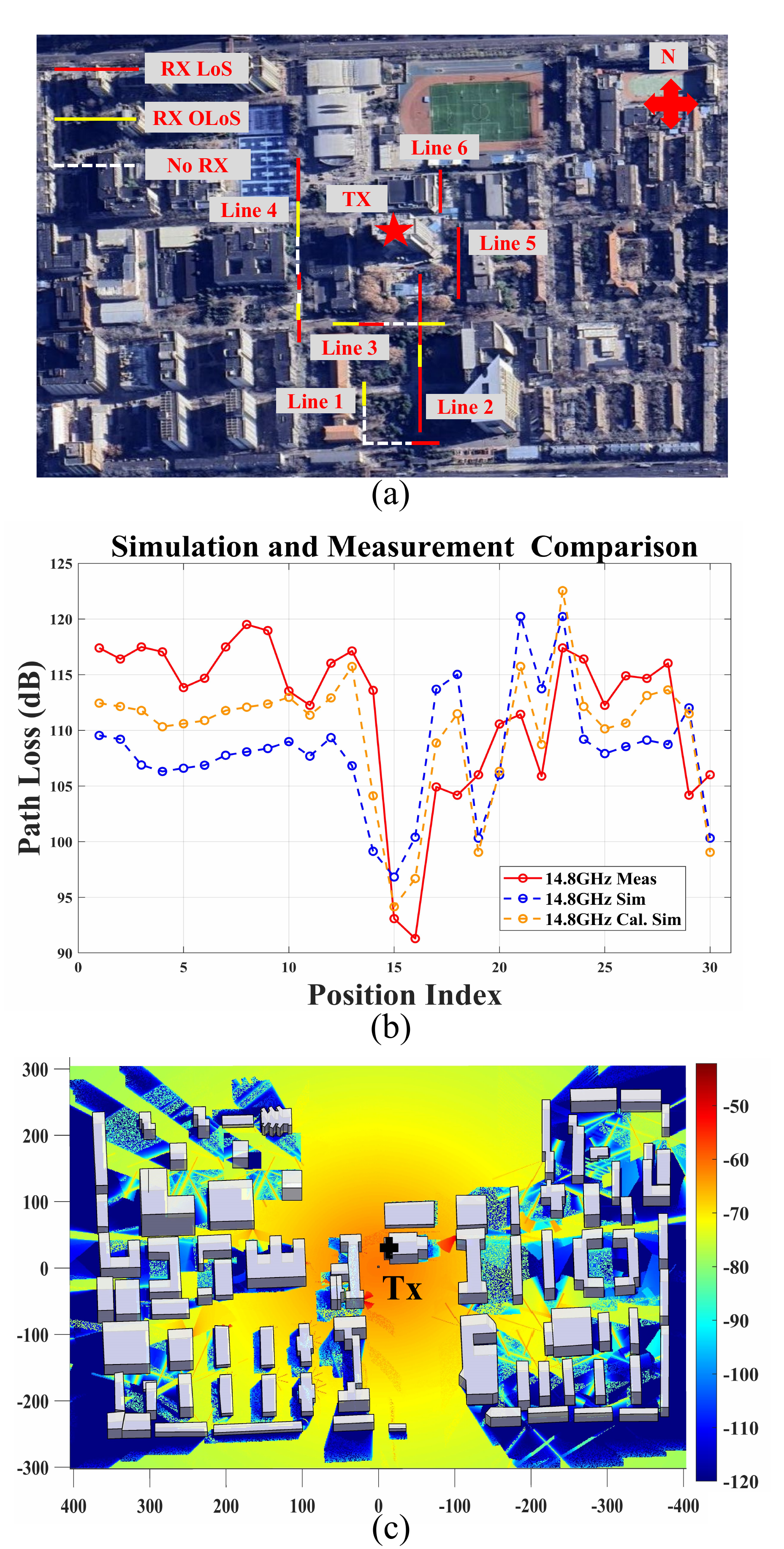}}
\caption{\textcolor{black}{Offline task validation in the campus scenario: (a) shows the measurement routes, (b) illustrates the comparison of PL errors before and after calibration, and (c) presents the prediction of the coverage capability for the actual base station.}}
\label{Offline task validation}
\end{figure}

Fig. 3(b) shows the PL error results for the 30 validation points before and after calibration. The PL error decreased from 7.7 dB to 4.5 dB, demonstrating the effectiveness of the calibration. Fig. 3(c) presents the coverage prediction of the MART-6G at 14.8 GHz after calibration. The base station can cover the vast majority of the area, with some coverage gaps at the edges and in building shadow areas. It is recommended to consider adding small base stations or using RIS at appropriate locations to address these issues.
\subsection{Online Task Validation}
For the online validation, we use the physical layer link construction task assigned by DTN, selecting a vehicle-to-vehicle (V2V) dynamic scenario in the Beijing Yizhuang autonomous driving demonstration zone. As shown in Fig. 3(a), the Tx and Rx are placed at a three-way intersection, with initial speeds of 36 km/h and 18 km/h, respectively. The Tx reaches the intersection first, causing the Rx to slow down and stop. After the Tx moves away, the Rx resumes motion. Channel measurements are conducted during this process, with  omnidirectional antennas for both Tx and Rx  at 6 GHz frequency band. Meanwhile, a 3D reconstruction of the surrounding road scene was created to support radio propagation calculations, as shown in Fig. 4(b).


\begin{figure*}[htbp]
\centerline{\includegraphics[scale=0.44]{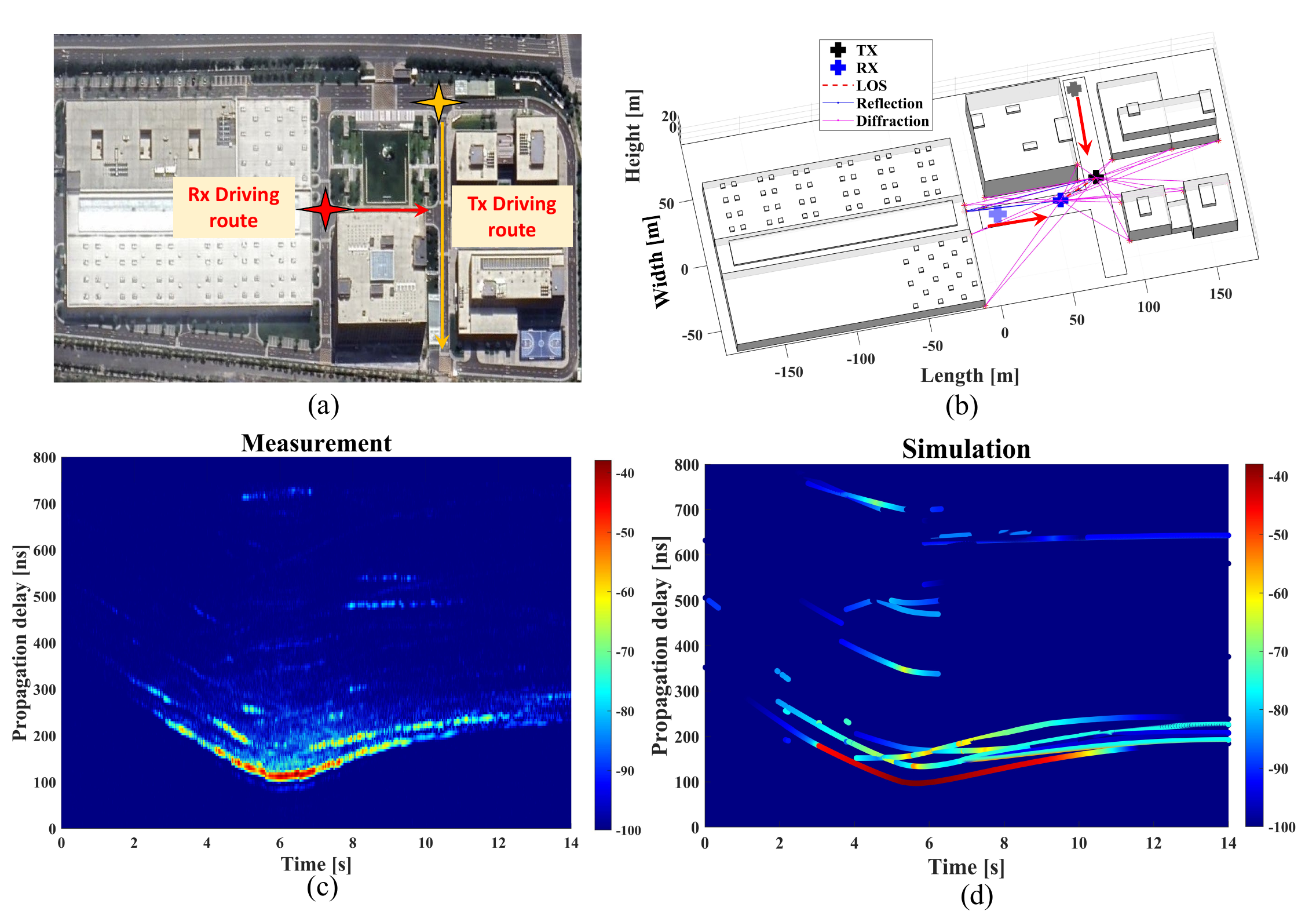}}
\caption{\textcolor{black}{Online task validation in the V2V scenario: (a) shows the vehicle route, (b) presents the radio propagation prediction, and (c) and (d) display the distribution of MPCs for the Rx based on measurements and simulations.}}
\label{Offline task validation}
\end{figure*}

Fig. 4(c) shows the distribution of MPCs based on measurement results, with the x-axis representing the Rx’s distance traveled and the y-axis representing the delay. When Rx moves for 3 seconds, the LOS path appears. At 6 seconds, the distance between the Rx and Tx is minimized, resulting in the minimum LOS delay. As the Tx moves further away, the LOS delay increases and the power decreases. Fig. 4(d) shows the MPC distribution output by the simulator, which closely matches the measurement results. To assess the accuracy of the twin channel, a channel similarity index (SI) is introduced\cite{15_yuanRTMIMO}. The SI ranges from 0 to 100\%, where 100\% represents perfect similarity and 0 represents maximum dissimilarity. The SI between the twin channel and the measurement channel is 86\%. In terms of computation speed, the simulator’s average update time is 70 ms, meeting the real-time requirements of the task.


	
	
	
	
	
	
	
	
	

\begin{table}[htbp]
\centering
\caption{Comparison between calibration and standard RT.}
\label{tab:rt-comparison}
\normalsize  
\begin{tabular}{@{}lcccc@{}}  
	\toprule
	\multirow{2}{*}{\textbf{Comparison}} & 
	\multicolumn{2}{c}{\textbf{PL }} & 
	\multicolumn{2}{c}{\textbf{DS }} \\

	\cmidrule(lr){2-3} \cmidrule(lr){4-5}
	& \textbf{RMSE} & \textbf{NMSE} & \textbf{RMSE} & \textbf{NMSE}  \\  
	
	\midrule
	Calibration RT & 4.99 dB & 0.002 & 5.7 ns & 0.005 \\
	
	Standard RT & 8.11 dB & 0.005 & 9.6 ns & 0.008 \\
	
	\cmidrule(lr){2-5}
	
	\multirow{4}{*}{ } & 
	\multicolumn{2}{c}{\makecell{\textbf{Online run time} \\ \textbf{(s/point)}}} & 
	\multicolumn{2}{c}{\makecell{\textbf{Offline run time} \\ \textbf{(s/1000 points)}}} \\
	
	\cmidrule(lr){2-5}
	
	Calibration RT & 
	\multicolumn{2}{c}{0.68} & 
	\multicolumn{2}{c}{249} \\
	
		\cmidrule(lr){2-5}
	
	Standard RT & 
	\multicolumn{2}{c}{0.07} & 
	\multicolumn{2}{c}{40} \\
	
	\bottomrule
\end{tabular}
\end{table}

To further demonstrate the performance of the MART-6G, Table \ref{tab:rt-comparison} presents the average results from large-scale tests conducted on the two scenarios mentioned above. After calibration, the simulator significantly improves accuracy, with the root mean square error (NMSE) for PL and DS decreasing from 0.005 to 0.002 and 0.008 to 0.005, respectively. At the same time, the calibration process results in longer simulation times. Therefore, for online tasks, we recommend using the standard RT, with an average simulation time of 0.07 s per point. For offline tasks, we suggest using the calibrated RT, with an average simulation time of 249 s for every 1000 points.

\section{Challenges and Future Opportunities}
Despite its promising potential as a key enabler for 6G DTN, the MART-6G is still in its early stages and faces several challenges that require further research. These challenges can be categorized into three main areas.

\textbf{Large-scale Model Validation and Calibration:} The accuracy of the MART-6G depends on how well its models and databases align with the real-world environment. A major challenge is how to efficiently and economically collect and process real-world measurements for calibrating the RT engine. Furthermore, efficient and precise calibration methods, such as those incorporating artificial intelligence, remain an area that requires further exploration.

\textbf{Integration with Existing Communication Systems:} For the effective deployment of the MART-6G, integrating it with existing communication systems is a critical consideration. The diversity of communication protocols, device architectures, and standardization requirements may pose compatibility and interoperability challenges.

\textbf{Hardware and Energy Consumption:} RT typically demands significant computational resources and energy. When applied at scale in 6G networks, this can lead to conflicts between hardware load and energy efficiency. Therefore, the primary challenge lies in achieving energy-efficient operation without compromising performance.

\section{Conclusion}

This paper proposes a MART-6G platform tailored for DTN tasks to generate accurate wireless channel with 6G features. We first examine the new requirements for 6G channel modeling and the challenges RT faced. Then, this paper details the role of RT in enabling DTN, and introduces the concepts and key features of online and offline twin tasks. A comprehensive overview of the MART-6G follows, covering its functional modules and innovative features. Simulation results show that the NMSE of PL and DS between real-world measurements and the MART-6G outputs are 0.002 and 0.005, respectively. In addition, online real-time channel predictions are achieved in less than 0.1 seconds with satisfactory accuracy. Finally, open issues related to RT-enabled DTN are discussed.

\section{Acknowledgement}
This work is supported by the National Natural Science Foundation of China (No. 62401084), the National Key R\&D Program of China (Grant No. 2023YFB2904805), and BUPTCMCC Joint Innovation Center.

\bibliographystyle{IEEEtran}
\bibliography{cite}

\begin{IEEEbiographynophoto}{LI YU (li.yu@bupt.edu.cn)}
is currently a postdoctoral research fellow with Beijing University of Posts and Telecommunications.
\end{IEEEbiographynophoto}

\begin{IEEEbiographynophoto}{YINGHE MIAO (yhmiao@bupt.edu.cn)}
is currently pursuing the master’s degree with Beijing University of Posts and Telecommunications. 
\end{IEEEbiographynophoto}

\begin{IEEEbiographynophoto}{JIANHUA ZHANG (jhzhang@bupt.edu.cn) }
is currently a professor with Beijing University of Posts and Telecommunications.
\end{IEEEbiographynophoto}

\begin{IEEEbiographynophoto}{SHAOYI LIU (sy\_liu@bupt.edu.cn)}
is currently pursuing the Ph.D. degree with Beijing University of Posts and Telecommunications. 
\end{IEEEbiographynophoto}

\begin{IEEEbiographynophoto}{YUXIANG ZHANG (zhangyx@bupt.edu.cn) }
is currently a associate researcher in Beijing University of Posts and Telecommunications.
\end{IEEEbiographynophoto}

\begin{IEEEbiographynophoto}{GUANGYI LIU (liuguangyi@chinamobile.com) }
is currently Fellow and 6G lead specialist, China Mobile Research Institute.
\end{IEEEbiographynophoto}

\end{document}